# Evidence of field induced interferroelectric transformation as the dominant driving mechanism for anomalous piezoresponse in morphotropic phase boundary piezoelectric system PbTiO$_3$-BiScO$_3$

Lalitha K.V.[1], Chris M. Fancher[2], Jacob L. Jones[2] and Rajeev Ranjan[1*]

[1]Department of Materials Engineering, Indian Institute of Science, Bangalore 560012, India

[2]Department of Materials Science and Engineering, North Carolina State University, Raleigh, NC 27695, USA

## Abstract

The contributory mechanisms associated with high piezoelectric response in piezoelectric ceramics have been examined by in-situ electric field dependent high energy synchro x-ray diffraction study. A comparative study of electric field induced lattice strain and the propensity for non-180$^o$ domain switching on two closeby compositions of a high performance piezoelectric alloy (1-x)PbTiO$_3$-(x)BiScO$_3$, one within the morphotropic phase boundary (MPB) region exhibiting d$_{33}$ of 425 pC/N and another just outside the MPB region exhibiting d$_{33}$ of 260 pC/N, unravelled that, inspite of the MPB specimen exhibiting considerably high piezoelectric response, its lattice strain and domain switching propensity is considerably less as compared to the non-MPB specimen. These new experimental observations contradict the commonly held view that anomalous piezoelectric response in MPB based piezoelectrics arise due to enhanced propensity for domain switching. Our results show the dominant mechanism contributing to the anomalous piezo-response is field induced interferroelectric transformation.

*rajeev@materials.iisc.ernet.in

Morphotropic phase boundary (MPB) ferroelctrics such as $Pb(Zr_x Ti_{1-x})O_3$, are widely as actuators, sensors and transducers by virtue of its exceptionally large piezoelectric response. The exceptional properties occurs at the MPB composition which exhibts coexistence of rhombohedral and tetragonal phases with spontaneous polarization along $[111]_c$ and $[001]_c$ directions, respecively [1, 2]. The first intuitive explanation for the large electromechanical response was attributed to the availability of large number of domain variants which was supposed to enable efficient poling of the specimen [3]. This idea received theoretical support from a Devonshire-Ginzburg-Landau based multiscale calculations which predicted enhanced domain switching in MPB systems [4]. The same scenario is considered to be applicable in single phase low symmetry ferroelectrics. In contrast, first principles [5, 6] and phenomenological free energy calculations [7-9] have shown a correlation between anisotropic flattening of free energy profile and polarization rotation, assisted by low symmetry phases, as the fundamental mechanism for enhanced piezoelectric response in ferroelectrics. It is generally agreed that there are three key contributing mechanisms with regard to the piezoresponse in MPB ferroelectrics: (i) lattice strain, (ii) domain wall dispalcement, and (iii) interferroelectric phase transformation. While several studies in the past have interpreted the results of field dependent diffraction experiments in the framework of polarization rotation theory [10-12], a competing view attributes the anomalous piezoelectric response in ferroelectric to enhanced density and mobility of domain-walls [13-18] near MPB. Raleigh law based phenomenological approach has generally been used to parameterize the domain wall and lattice contributions in ferroelectrics [19-24]. In-situ electric field diffraction experiment enables direct estimation of domain switching and lattice strain in ferroelectrics [24-30]. Though, in principle, such a technique can also ascertain the occurrence of field-induced interferroelectric transformation, if any, the complexity associated with preferred orientation makes interpretation of the nature of interferroelectric transformation ambiguous. As a result, even in cases where inter-ferroelectric transformation has been recognized [26, 30], it has not been possible to ascertain the relative role of the three contributing mechanisms stated above, vis-à-vis the anomalous piezoelectric response. In the present work, we have addressed this important problem by way of comparative analysis of the lattice strain and domain switching propensity on two compositions a piezoelectric alloy $(1-x)PbTiO_3-(x)BiScO_3$ (BSPT), one showing MPB (x = 0.3725) and high $d_{33}$ (425 pC/N) and a non-MPB composition (x = 0.40) exhibiting signficanlty less $d_{33}$ (260 pC/N) [31, 32]. Contrary to the common perception, our results demonstrate that domain mobility and lattice strain in the MPB specimen is nearly half of the non-MPB specimen. The predominant contributing mechanism for the anomalous piezoelectric response in the high performace MPB ferroelectrics is instead the inter-ferroelectric transition.

*In situ* electric field high energy x-ray diffraction experiment was carried out at the Advanced Photon Source at Argonne National Laboratory in transmission geometry, which ensured that the measured diffraction data probes the bulk response of the specimen. A monochromatic beam of wavelength 0.11165Å and size 500 µm x 500 µm was used for the diffraction experiments. The disk-shaped ceramic samples were cut to 10 x 1 x 1 mm³ (l*b*t) dimensions and electric field was applied across 10 mm x 1 mm faces of the sample. Electroding was done using silver paste across the surface to which electric field was applied. The data was collected as a 2D image, wherein, the circular Debye rings correspond to different *hkl* diffracted beams. The schematic details of the in-situ diffraction geometry is shown in Fig. 1. Ceria ($CeO_2$) was used as the standard to calibrate the sample to detector distance. The diffraction images were divided into 24 azimuthal sectors of 15º widths, with the azimuthal sector most closely oriented to the direction of applied electric field defined as $\psi = 0º$. The diffracted intensities as a .3725 (MPB) and $x$ = 0.40 (monoclinic). The *in situ* electric field dependent study is generally carried out by applying a triangular bipolar wave form. The field was varied in incremental steps of 0.5 kV/mm and the diffraction pattern recorded. The amplitude of the field was 2.5 kv/mm. Piezoelectric strain of the ceramic specimens were measured using Radiant Premier Precision II set up with MTI photonic sensor. Direct piezoelectric coefficient was measured with a Berlincourt based piezometer (Piezotest PM 300).

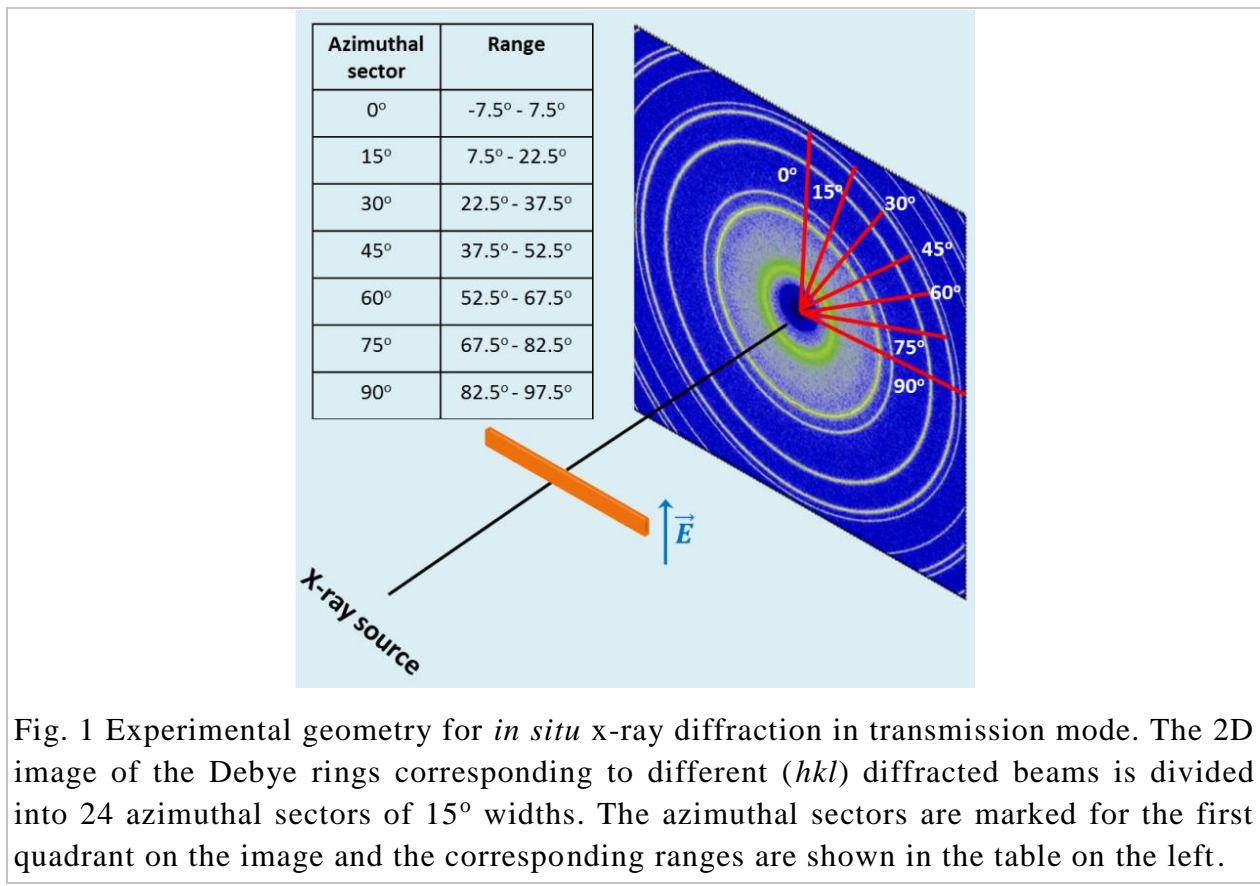

Fig. 1 Experimental geometry for *in situ* x-ray diffraction in transmission mode. The 2D image of the Debye rings corresponding to different (*hkl*) diffracted beams is divided into 24 azimuthal sectors of 15º widths. The azimuthal sectors are marked for the first quadrant on the image and the corresponding ranges are shown in the table on the left.

Fig. 2 shows electric field dependent longitudinal strain of the non-MPB and the MPB compositions, respectively. Evidently, at any given field the magnitude of strain of non-MPB composition is less than that of the MPB composition. For example at E = 2.5 kV/mm, the positive strains are 0.11 % and 0.21 % for the non-MPB and the MPB composition, respectively. This is also consistent with the considerably large $d_{33}$ (as measured by Berlincourt based piezometer) of the MPB composition as compared to the non-MPB composition, mentioned above.

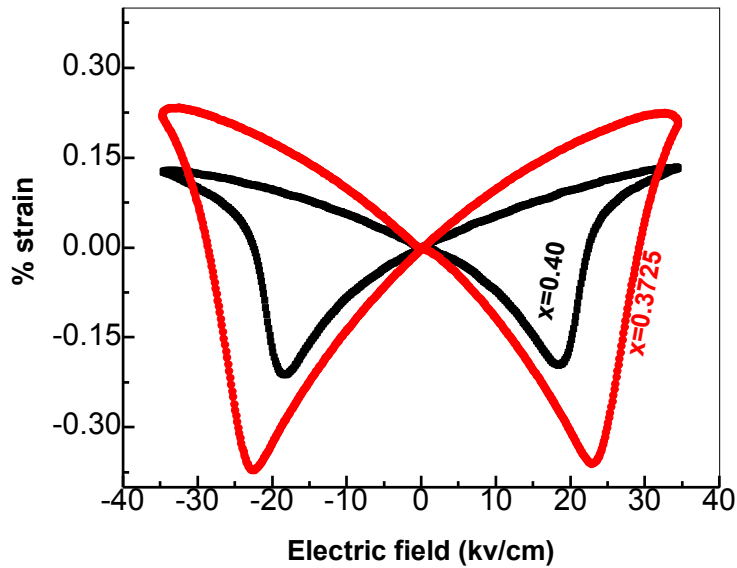

Figure 2: Strain – field response of (x)BS-(1-x)PT for x=0.40 (non MPB) and x=0.3725 (MPB) measured at 1 Hz.

Similar to PZT, the MPB composition of this alloy has been characterized as a coexistence of rhombohedral and tetragonal phase [33]. Though there is no distinct signature of deviation from the rhombohedral phase in the XRD pattern of these specimens, high resolution diffraction data has been reported to fit better with a monoclinic (Cm) phase [34-36]. The interpretation of the diffraction data presented in this paper, however, does not require consideration of monoclinic symmetry in both the compositions. Hence, for sake of simplicity, we would consider the non-MPB composition (x=0.40) and the MPB composition (x=0.3725) as exhibiting rhombohedral and rhombohedral + tetragonal phases, respectively, as was reported earlier by Eitel et al [33]. On application of electric field, the domains tend to reorient along the field direction. This phenomenon manifests as relative change in the intensity of the symmetry related Bagg peaks. For a rhombohedral phase with polarization along the [111] direction, electric field would increase the intensity of the $(111)_R$ rhombohedral peak at the expense of the intensity of the $(11\text{-}1)_R$ rhombohedral peak. Similarly, for the tetragonal phase whose spontaneous polarization is along $[001]_T$, the intensity of the $(00l)_T$ tetragonal peak would increase at the expense of the intensity of the $(h00)_T$ peak. Here, we have assumed that the indices of the direction is normal to the plane with the same indices. This assumption is valid for small rhombohedral/tetragonal distortion from the cubic structure, as is usually the case. In rhombohedral ferroelectric perovskites, the volume fraction of [111] domains which has been reoriented through application of electric field is given by [25, 26]

$$\eta_{111} = \frac{\dfrac{I_{111}}{I'_{111}}}{\dfrac{I_{111}}{I'_{111}} + 3\dfrac{I_{11\bar{1}}}{I'_{11\bar{1}}}} - \frac{1}{4}$$

where, $I_{111}$ and $I_{11\bar{1}}$ are the integrated intensities of the (111) and $(11\bar{1})$ reflections respectively in the poled state of the material exhibiting preferred orientation of non-180° domains. $I'_{111}$ and $I'_{11\bar{1}}$ are the integrated intensities of the (111) and $(11\bar{1})$ reflections respectively in the unpoled state of the material with random domain configuration. Similarly, for the tetragonal ferroelectric perovskite, the volume fraction field induced domain reorientation is given by

$$\eta_{002} = \frac{\dfrac{I_{002}}{I'_{002}}}{\dfrac{I_{002}}{I'_{002}} + 2\dfrac{I_{200}}{I'_{200}}} - \frac{1}{3}$$

where, $I_{002}$ and $I_{200}$ are the integrated intensities of the $(002)_T$ and $\{200\}_T$ reflections respectively in the presence of field. $I'_{002}$ and $I'_{200}$ are the integrated intensities of the $(002)_T$ and $\{200\}_T$ reflections respectively before application of the field. Apart from estimation of the domain reorientation, the shift in the peak positions with electric field can be used to measure lattice strain [25, 26]

$$\varepsilon_{hkl} = \frac{d_{hkl}(E) - d_{hkl}(0)}{d_{hkl}(0)}$$

Fig. 3 shows the change in the shapes of the $\{111\}_c$ and $\{200\}_c$ pseudocubic Bragg profiles with electric field of both the specimens. For the non-MPB composition, the two distinct noticeable changes are (i) reversal in the intensity ratio of the $(111)_R$ and $(11\text{-}1)_R$ reflections at high fields, and (ii) shift in Bragg peak position of the $(200)_R$ peak suggesting significant lattice strain in the non-polar direction. In fact both these quantities depend on the orientation of the plane normal with respect to the electric field ($\psi$), as shown in Fig. 4, with their maximum values along the longitudinal direction, i.e. $\psi = 0^o$. In this paper, we will therefore concern ourselves with values of $\eta$ and $\varepsilon$ corresponding to $\psi = 0^o$.

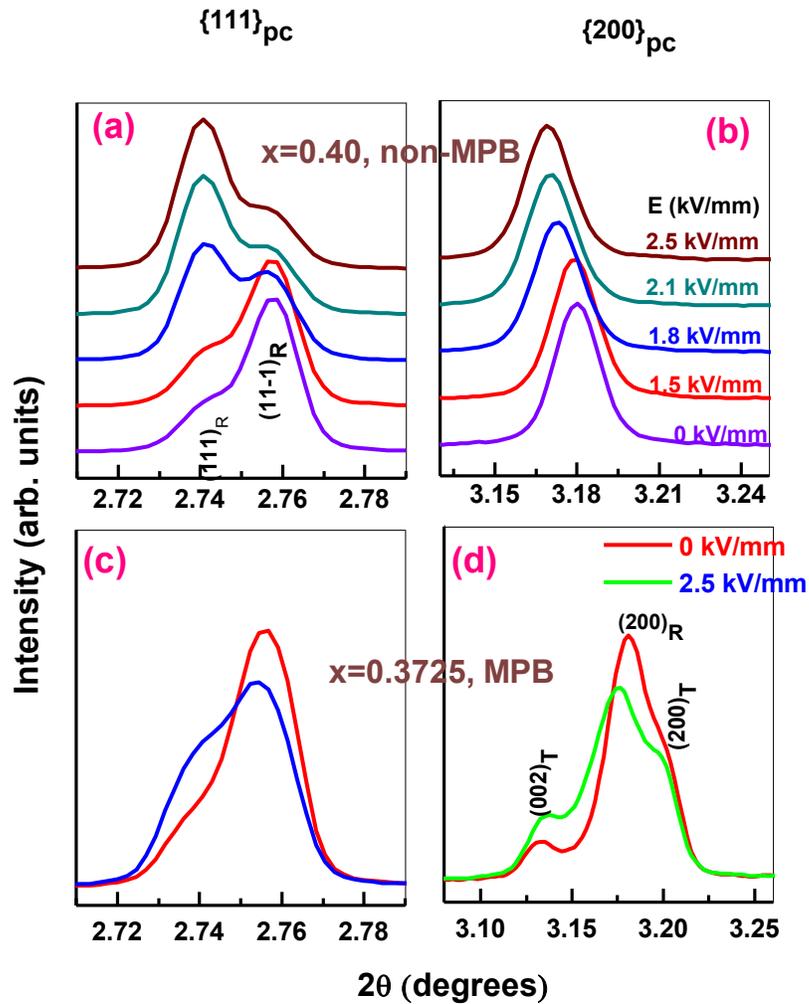

Figure 3. X-ray diffraction spectra of $\{111\}_{pc}$ and $\{200\}_{pc}$ Bragg profiles of BS40 showing shift in peak positions as a function of electric field for $\psi = 0^o$ **(a-b)** and orientation with respect to the direction of electric field **(c-d)** for an applied electric field of 2.5kV/mm.

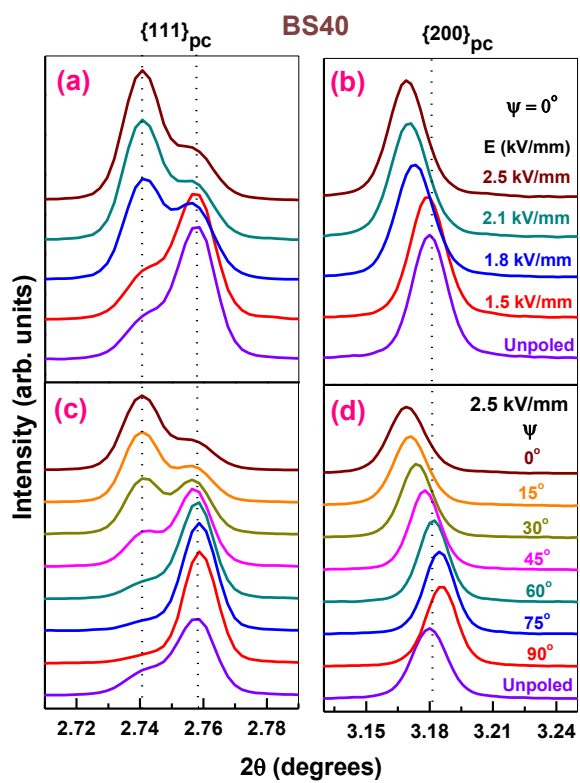

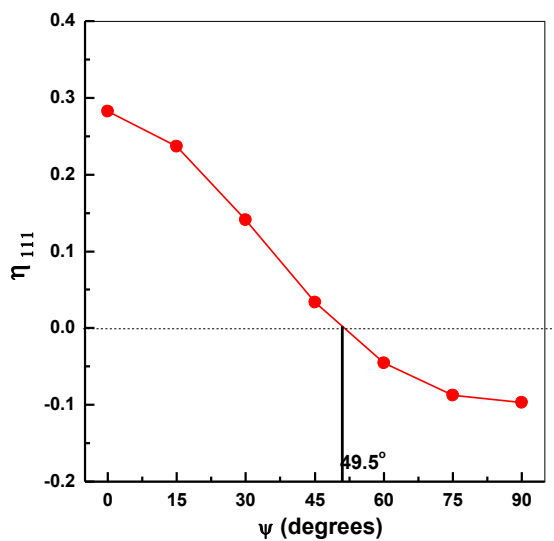

**Fig. 4**. X-ray diffraction spectra of $\{111\}_{pc}$ and $\{200\}_{pc}$ Bragg profiles of the non-MPB composition x=0.400 showing shift in peak positions as a function of electric field for $\psi$ = 0° **(a-b)** and orientation with respect to the direction of electric field **(c-d)** for an applied electric field of 2.5kV/mm. Variation of $\eta_{111}$ of with azimuth angle $\psi$ for applied field of 2.5 kV/mm is shown in the bottom panel.

Fig. 5 shows the field dependence of $\eta_{111}$ and $\varepsilon_{200}$ when the magnitude of the field was increased from 0 to 2.5 kV/mm. Both the quantities show exactly the same trend with electric field, thereby proving the existence of a strong coupling between lattice strain and reorientation of non-180° domains, a phenomenon generally attributed to stress field generated in the polycrystalline specimen due to shape change of the grains during domain reorientation[25, 26, 37].

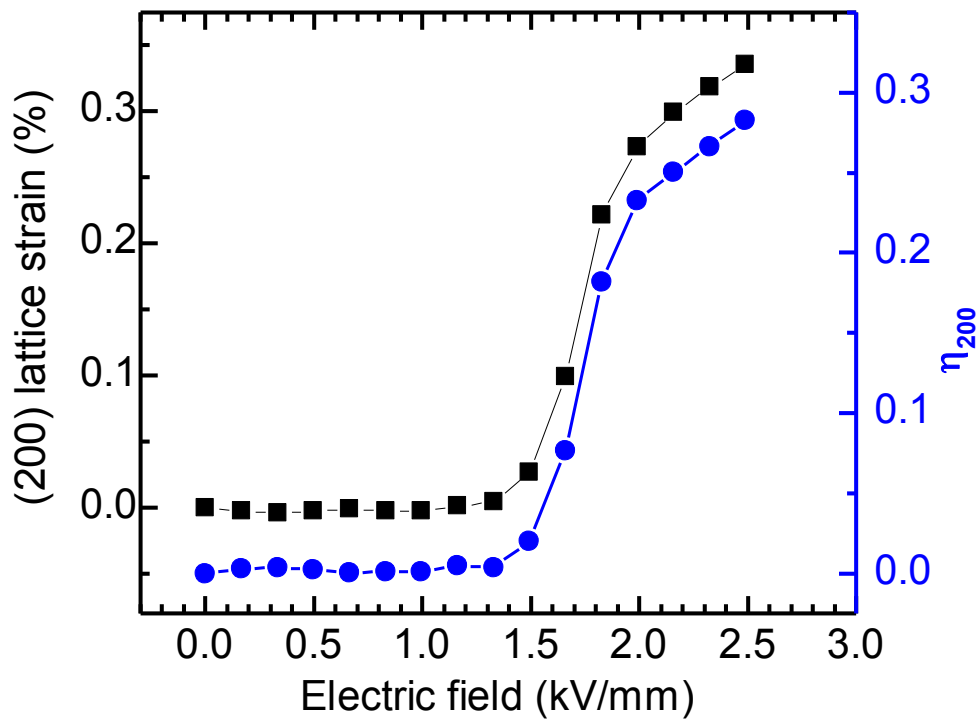

Fig. 5 Field dependence of $\varepsilon_{200}$ and $\eta_{200}$ of the non-MPB composition.

It may be pointed out that, compared to $\varepsilon_{200}$ the lattice strain along the polar direction $\varepsilon_{111}$ is nearly five times small (at 2.5 kV/mm $\varepsilon_{200} = 0.32$ and $\varepsilon_{111} = 0.07$ %). A similar result by Guo et al. was interpreted as proof of the polarization rotation away from the polar direction [10]. This interpretation was rationalized with the theoretical [38] and experimental [39] reports that predicted large piezoelectric response along non-polar directions in single crystals. As per the equation mentioned above, a complete reorientation of all the [111]$_R$ domains would result in $\eta_{111} = 0.75$. However at 2.5 kV/mm, which is well above the coercive field, $\eta_{111} = 0.28$, implying 37 % domain reorientation in the non-MPB specimen. In a separate experiment we also recorded data at 3 and 4.5 kV/mm. At 4.5 kV/mm, which is far above the coercive field (1.8 kV/mm), the domain reorientation increased to ~ 50%. Hence complete domain reorientation is not likely to happen at any realizable field. Similar experiment was performed on the MPB composition (x=0.3725). For this composition, the $\{111\}_c$ and $\{200\}_c$ pseudocubic profiles appear as doublet and triplet, respectively. Rietveld analysis of the diffraction pattern corresponding to the unpoled state (i.e. before switching on the field) revealed that the $(111)_T$ peak of the tetragonal phase overlaps severely with $(11\text{-}1)_R$ profile of the rhombohedral phase, Fig. 6.

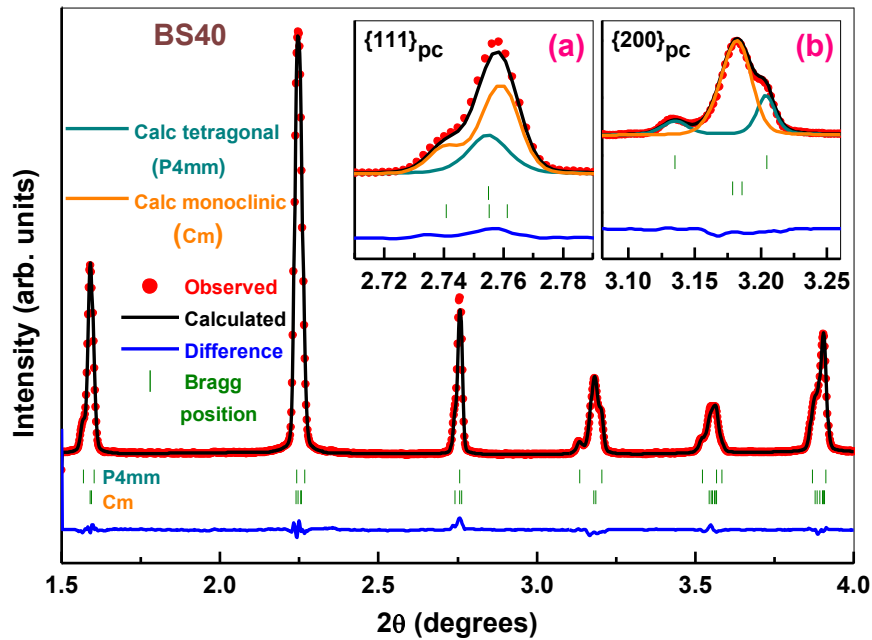

**Fig. 6:** Rietveld fit of high energy x-ray diffraction pattern of the MPB composition before switching on the field. In accordance with the literature the fitting was carried out with the tetragonal (T) + monoclinic (M) two phase model. The insets shows magnified view of the (a) $\{111\}_{pc}$ and (b) $\{200\}_{pc}$ profiles. The monoclinic phase was considered here instead of rhombohedral for better fitting of the data. In the analysis related to domain reorientation/switching in the text we have considered rhombohedral phase.

  The triplet in the $\{200\}_c$ has two peaks on the extreme $(002)_T$ and $(200)_T$ corresponding to the tetragonal phase and the one in the middle corresponding to the rhombohedral phase.  Since the $(002)_T$ and $(200)_T$ tetragonal peaks are reasonably separate from the rhombohedral $(200)_R$ peak, it enabled us to determine $\eta_{002}$ as a function of field (Fig. 7a).

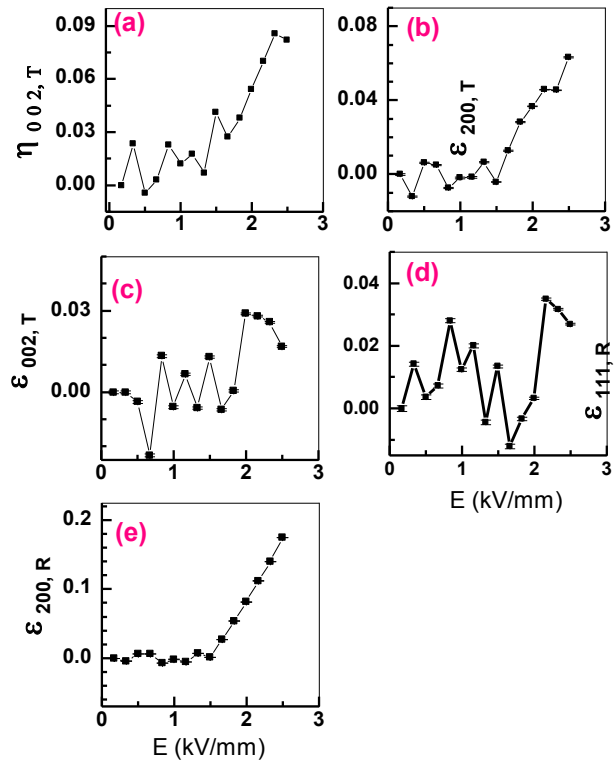

Fig. 7 Electric field dependence of tetragonal domain switching in the MPB composition. T and R in the subscripts refer to parameters of the tetragonal and rhombohedral phases, respectively.

The maximum value of this quantity was obtained as 0.09 at 2.5 kV/mm.  On the other hand, the severe overlap of Bragg peaks $(11\text{-}1)_R$ and $(111)_T$ corresponding to the rhombohedral and tetragonal phases precluded estimation of $\eta_{111}$. However, to get a qualitative feeling with regard to the propensity of 111 domain reorientation in the MPB composition vis-à-vis the non MPB composition, we compared the $\dfrac{I_{111}}{I'_{111}}$ as a function of electric field, since this ratio is also an indicator of the same phenomenon. This ratio was found to be 2.5 for the non-MPB composition and 1.6 for the MPB composition (Fig. 8).

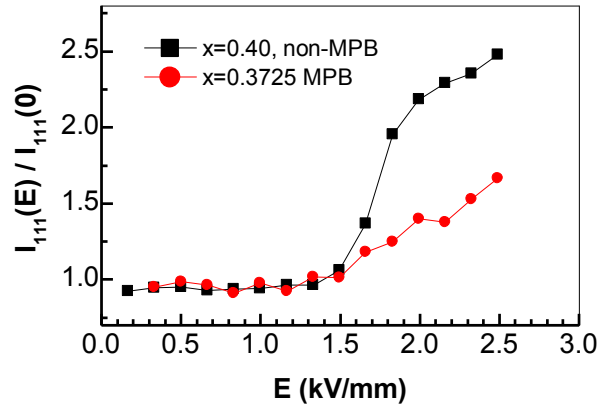

Fig. 8 Electric field dependence of the intensity of the $(111)_R$ peak, scaled with respect to the intensity at zero field, of the MPB and non-MPB composition.

The lattice strains obtained from the shift in the peak positions of non-overlapping Bragg peaks of the two phases are shown in Fig. 7b-7e. The largest lattice strain for the MPB composition occurs in $(200)_R$ of the rhombohedral phase. At 2.5 kV/mm this strain is 0.18 %, which is more than five times larger than the lattice strains in the coexisting tetragonal phase (<0.05% at 2.5 kV/mm). Most remarkably, $\varepsilon_{200, R}$ for the MPB composition is half the value of the non-MPB composition at 2.5 kV/mm (compare Fig 5 and Fig. 7e). Thus the specimen exhibiting nearly twice the piezoresponse shows nearly half the propensity for domain switching as well as lattice strain. This result is in complete disagreement with the commonly held view that the MPB compositions exhibit a greater propensity of domain switching and lattice strain [13-18].

It is important to highlight that while for the non-MPB composition there is complete reversal of the the intensities of $(111)_R$ and $(11\text{-}1)_R$ above the coercive field (Fig. 4a), for the MPB system (Fig 4c) the intensity at $(11\text{-}1)_R$ position remain larger than the intensity at the $(111)_R$ position at all fields. This apparently contrasting feature can be understood if we bear in mind that the tetragonal $(111)_T$ peak overlaps severely with the $(11\text{-}1)_R$ peak (Fig. 6). The enhanced intensity near the $(11\text{-}1)_R$ position therefore has dominant contribution from increased fraction of the tetragonal phase at high fields. We have noted that intensity at this position changes even at subcoercive fields, thereby proving the occurrence of field induced interferroelectric transformation even at low fields [30].  In view of these findings, the field induced structural transformation reported earlier by Lalitha et al [31] assumes great significance. Using ex-situ based approach, it was demonstrated that the composition (x=0.3725) exhibiting highest $d_{33}$ shows maximum field induced rhombohedral to tetragonal transformation (~20 %). This, in conjuyunction with the fact that domain wall mobility is considerably reduced for this composition as compared  to the non-MPB composition, categorically proves that it is the field-induced interferroelectric transformation, and not the domain wall mobility that causes anomalous piezoelectric response in the MPB composition. Using the same ex-situ based technique a close correlationship between high piezoelectric response and field induced interferroelectric transformation has also been reported in pure and dilutely modified $BaTiO_3$ [40, 41] and $Na_{1/2}Bi_{1/2}TiO_3$ [42] based systems. Our results suggests that anomalous properites in all ferroelectric alloys exhibiting large piezoelectric response is primarily due to enhanced propensity for interferroelectric transformation. Since the symptom of field driven transformation could not be seen in the in-situ experiment even for field as large as 4 kV/mm, the possibility of interferroelectric transformation is ruled out for the non-MPB composition. In a recent work, the equivalence of stress and electric field with regard to the irreversible structural changes has also been demonstrated for the MPB compositions [32, 43]. Keeping in view the intimate relationship between structure and piezoelectric properties, the anomalous piezoelectric response measured through direct piezoelectric effect (with stress as stimulus) and  also by converse piezoelectric effect (with electic field as stimulus), implies that same interferroelectric transformation plays  as the primary governing mechanism whether stress or electric field is used as the stimulus to measure the piezoelectric response. For single phase ferroeletrics, the applied electric energy is used in the displacement of domain walls and to strain the lattice. The fact that for the MPB composition both these quantities are considerably less than in the non-MPB composition, proves that the electric energy in the MPB specimen is primarily used in effecting interferroelectric transformation, and less in moving the domain wall within a given ferroelectric phase. Our results therefore prove that the piezoelectric response in the MPB system arise from a complex yet cooperative mechanism of interferroelectric transformation, domain switching and lattice strain, with interferroelectric transformation as the primary driving mechanism.

In conclusion, a comparative in-situ electric field dependent high energy synchrotron x-ray diffraction study on a MPB and a closeby non-MPB composition of  the ferroelectric alloy $BiScO_3$-$PbTiO_3$ revealed that the propensity of non-180 [0] ferroelectric-ferroelastic domain switching and the coupled elastic lattice strain is considerably low in the MPB composition in comparison to the non-MPB composition. This new finding contradicts the commonly held view that the primary mechanism associated with the anomalous piezoelectric response of MPB piezoelectrics is high domain wall mobility.

Our results rather demonstrate the primary mechanism to be field induced interferroelectric transformation. Domain switching and the coupled-lattice-strain is rather a secondary mechanism which occur cooperatively with the field-induced interferroelectric transformation to determine the overall piezoelectric response in MPB piezoelectric. These results offer a new fundamental perspective with regard to the mechanisms in MPB based high performance piezoceramics and is likely to stimulate further work on similar lines.